\documentclass[graybox]{svmult}

\usepackage{type1cm}        
\usepackage{makeidx}         
\usepackage{graphicx}        
\usepackage{multicol}        
\usepackage[bottom]{footmisc}

\usepackage{newtxtext}       %
\usepackage[varvw]{newtxmath}       

\usepackage{xurl}
\usepackage[colorlinks,allcolors=blue]{hyperref}
\usepackage[authoryear]{natbib}

\makeindex             

\newcommand{\eps}{\varepsilon}
\newcommand{\what}[1]{\widehat{#1}}
\newcommand{\Ee}{\mathsf{E}}
\newcommand{\op}{o_{\mathsf{P}}}
\newcommand{\R}{\mathbb{R}}
\newcommand{\Pto}{\stackrel{\mathsf{P}}\to}

\begin{document}

\title*{Specification Tests for the Error--Law in Vector Multiplicative Errors Models}
 
 \author{\v{S}\'{a}rka Hudecov\'{a} 
 and\\ Simos G. Meintanis}

 \institute{\v{S}\'{a}rka Hudecov\'{a}\at Department of Probability and Mathematical Statistics, Faculty of Mathematics and Physics, Charles University, Czech Republic \email{hudecova@karlin.mff.cuni.cz}
\and Simos G. Meintanis \at Department of Economics, National and Kapodistrian University
of Athens, Greece and  Pure and Applied Analytics, North--West University, Potchefstroom, South Africa}
 
%
%
%
%
%
%
%


\maketitle

\abstract*{We suggest specification tests for the error distribution in vector multiplicative error models (vMEM). The test statistic is formulated as a weighted integrated distance between the parametric estimator of the Laplace transform of the null distribution and its empirical counterpart computed from the residuals. Asymptotic results are obtained under both the null and alternative hypotheses.
If the Laplace transform of the null distribution is not available in closed form, we propose a test statistic that uses independent artificial samples generated from the distribution under test, possibly with estimated parameters. The test statistic compares the empirical Laplace transforms of the residuals and the artificial errors using a similar weighted integrated distance. Bootstrap resampling is used to approximate the critical values of the test.
The finite--sample performance of the two testing procedures is compared in a Monte Carlo simulation study.}

\abstract{We suggest specification tests for the error distribution in vector multiplicative error models (vMEM). The test statistic is formulated as a weighted integrated distance between the parametric estimator of the Laplace transform of the null distribution and its empirical counterpart computed from the residuals. Asymptotic results are obtained under both the null and alternative hypotheses.
If the Laplace transform of the null distribution is not available in closed form, we propose a test statistic that uses independent artificial samples generated from the distribution under test, possibly with estimated parameters. The test statistic compares the empirical Laplace transforms of the residuals and the artificial errors using a similar weighted integrated distance. Bootstrap resampling is used to approximate the critical values of the test.
The finite--sample performance of the two testing procedures is compared in a Monte Carlo simulation study.
}


 \section{Introduction}\label{sec_0}

Vector Multiplicative Error Models (vMEM) are a class of econometric models suitable for modeling of  non-negative multivariate time series, particularly in finance and economics   \citep{Cip13}. They are a multivariate extension of  the univariate  MEM, introduced by \cite{Engle2002}. In vMEM,  the current observation is expressed as the Hadamard product of the conditional expectation and a non-negative vector error term  representing unpredictable
news. 
In various applications, it is desirable to predict not only the conditional mean of the process but also its entire conditional distribution. This is crucial in situations where uncertainty quantification and risk assessment are important, such as volatility forecasting, liquidity modeling, and market microstructure analysis.
The construction of such probability forecasts requires the correct specification of the conditional mean, along with an appropriate specification of the error distribution. 

Several methods have been proposed in the literature for testing a given parametric specification for the conditional mean, 
 \citep{Ng,Sango,Ke,MNH}. However, 
to the best of our knowledge, the existing literature related to the problem of testing the error distribution  is  restricted to univariate MEM; the interested reader is referred to \cite{MMO}, \cite{Perera},  \cite{LO2015}, and \cite{FG2005}.  

This paper proposes a test for a hypothesis that the error distribution belongs to a given parametric class of distributions. We suggest a test statistic that measures the distance between the parametric 
 estimator of the Laplace transforms (LT) under the null hypothesis and the empirical LT of the residuals. Since its limiting distribution depends on unknown nuisance parameters,  a parametric bootstrap method is recommended for the implementation. In addition, since the LT of the null distribution is sometimes not available in a closed form, we also propose a method that makes use of an artificial sample generated from the null distribution with estimated parameters. 
 The corresponding test statistic then reflects the distance between the empirical LT of the residuals and the artificial sample.

The Laplace transform of a distribution is closely related to its Fourier transform, a tool that  have been successively used in the work of Professor Marie Hu\v{s}kov\'{a}, in a collaboration with Simos G. Meintanis and other coauthors, to test various specification hypotheses. Particularly, a change point detection \citep{HuCH1,HuCH2}, independence testing in various regression setups \citep{HuIND-hetreg,HuIND-SA} and for functional data \citep{HuIND-F2,HuIND-F1}, and specification tests for error distributions in various setups \citep{HuGOF-2,HuGOF-1,HuGOF-4,HuGOF-5,HuGOF-3}, to name some of them.

The paper is outlined as follows. 
Section~\ref{sec_1} presents the  model and the considered hypothesis.  The LT based test statistic is introduced in Section~\ref{sec_test}. Its limiting distribution is provided in Section~\ref{sec_as}, while the bootstrap test is described in Section~\ref{sec_boot1}. The variant based on the artificial sample is proposed in Section~\ref{sec_artif}. 
The finite--sample properties of both methods are compared in  a Monte Carlo simulation study in Section~\ref{sec_6}. 
The paper concludes with Section \ref{sec_7}.

\section{Considered model and the null hypothesis}\label{sec_1}

Let $\{x_t\}$ be a $d$-dimensional time series of non-negative random variables, $d\geq 1$, $x_t=(x_{t,1},\ldots,x_{t,d})^\top$, and define $\mathcal{F}_{t} = \sigma\{x_s, s\leq t\}$ the   information set  available at time $t$. 
 Consider the vector multiplicative error model (vMEM)   specified as 
\begin{equation}
\label{vMEM}
x_t=\mu_t \odot \varepsilon_t, \quad \mu_t = \mu(\vartheta;{\cal{F}}_{t-1}),
\end{equation}
where $\odot$ is the Hadamard (element-wise) product,  $\{\varepsilon_t\}$ 
are independent copies of a non-negative $d$-dimensional random  vector $\varepsilon$ such that $\mathsf E(\varepsilon)=\boldsymbol{1}_d:=(1,\ldots,1)^\top$, and $\mu$ is some function that specifies the dependence of 
 $\mu_t=\mathsf E(x_t|{\cal{F}}_{t-1})$  on the past $\mathcal{F}_{t-1}$ via some unknown  parameter vector $\vartheta \in \mathbb R^K$ for some $K\geq 1$. 
The far most popular choice of $\mu$ corresponds to the GARCH--type vMEM$(p,q)$ model with  
\begin{equation}\label{GARCH}
\mu_{t}=\alpha_0+\sum_{j=1}^p A_j x_{t-j}+\sum_{k=1}^q B_j \mu_{t-k},
\end{equation}
where $\vartheta$ is a vector of dimension $d+d^2(p+q)$, incorporating the $(d\times 1)$ vector $\alpha_0$  of positive elements, and the $(d\times d)$ matrices $A_j, \ j=1,\ldots,p$, and $B_k, \ k=1,\ldots,q$,  with nonnegative elements. 
Several instances of this model have been found useful in financial applications; see for instance \cite{Manganelli}, \cite{Engle}, \cite{Cip06}, and \cite{Cip13}, among others.   
For some recent generalizations see, for instance, \cite{Cip22}.  In the following we assume that the model in \eqref{vMEM} is specified correctly.

In this paper we address the problem of correct specification for the  distribution of the innovations $\varepsilon_t$, figuring in \eqref{vMEM}.  Let ${\cal{G}}=\{G_\varphi; \ \varphi\in \Phi\}$  be a specified family of distributions on $\R_d^+$ indexed   by a parameter $\varphi \in \Phi$, $\Phi\subseteq \mathbb R^r$, $r\geq 1$, and let $F_{\eps}$ be the distribution function of $\eps$. 
The aim is to test the null hypothesis 
\begin{equation} \label{null}
    \mathcal{H}_0:  F_{\eps} \in  {\cal{G}}. 
\end{equation}
In other words, we wish to test whether $F_{\eps} = G_{\varphi}$ for some  $\varphi \in \Phi$, against a general alternative. 
A possible interesting parametric specification $\mathcal{G}$ for $\varepsilon$ are the multivariate Gamma and the multivariate log-Normal distribution; see \cite{Cip17}, \cite{Engle} and \cite{Cip06}.  Copula--based specifications are also included, provided that the copula parameter $\varphi$ is fixed or can be estimated from the data; see \cite{Cip17}. 

%
%

\section{Test statistic}\label{sec_test}

Let  the data $x_1, \ldots, x_T$  come from a strictly stationary solution of \eqref{vMEM}. Since the true innovations $\{\eps_t\}$ are unobserved, the test of $\mathcal{H}_0$ needs to be based on some suitably defined residuals. 
Let $\widehat\vartheta_T$ be an estimator of the parameter vector $\vartheta$ constructed from the data, and let the estimated conditional mean
$\widehat \mu_t$, $t=1,\dots,T$,   be computed recursively using some initial values. For instance, for the GARCH-type vMEM in \eqref{GARCH}, one has
\[
\widehat{\mu}_{t}=\what{\alpha}_{T,0}+\sum_{j=1}^p \what{A}_{T,j} x_{t-j}+\sum_{k=1}^q \what{B}_{T,k} \widehat{\mu}_{t-j}, \quad t=1,\dots,T,
\]
for estimates $\what{\alpha}_{T,0}, \what{A}_{T,j},\what{B}_{T,k}$ of $\alpha_0, A_j, B_k$, respectively, $j=1,\dots,p$, $k=1,\dots,q$, and  
for some specified initial values for $\widehat \mu_t$ and $x_t$ for $t\leq 0$. Generally, we can write that $ \widehat \mu_t = \mu(\what{\vartheta}_T, \widetilde{\mathcal{F}}_{t-1})$, where 
$\widetilde{\mathcal{F}}_t=\sigma\{x_s, 1\leq s\leq t\}$ is the finite past. 
Define 
\begin{equation}\label{RES}
\widehat \varepsilon_t=x_t\oslash \widehat \mu_t, \ t=1,\ldots,T,
\end{equation}
where the symbol $\oslash$ denotes elementwise division of two vectors of the same dimension.
A reasonable approach to testing the null hypothesis $\mathcal{H}_0$ is based on a comparison of the empirical version of  the Laplace transform (LT) of $F_{\eps}$ computed from the residuals $\{\widehat{\eps}_t\}_{t=1}^T$ with a parametric estimator 
 constructed under the null hypothesis $\mathcal{H}_0$. 
 Remark that the LT has been successively used for goodness-of-fit testing in various setups, see, e.g.,  \cite{henze} for a test of exponentiality and \cite{MMO} for its application to univariate MEM.

For $\mathbb R_+^d:=(0,\infty)^d$ let
\[
L_{\varphi}(u)= \int_{\mathbb R_+^d} \mathrm{e}^{-u^\top y} \  \mathrm{d} G_{\varphi}(y), \quad  
u\in \mathbb R_+^d,
\]
be  the LT of $G_{\varphi}\in\mathcal{G}$ and
\begin{equation} \label{ELT}
\widehat{L}_T(u)= \frac{1}{T}\sum_{t=1}^T \mathrm{e}^{- u^\top \widehat \varepsilon_t}
\end{equation}
the empirical LT of the residuals $\{\widehat \varepsilon_t\}_{t=1}^T$. Assume that  $\widehat \varphi_T$ is an estimator of $\varphi$ constructed under $\mathcal{H}_0$. 
We propose  the test statistic   
\begin{equation} \label{test0}
S_{T,W}= T \int_{\mathbb R_+^d} \big[\widehat{L}_T(u)-L_{\widehat \varphi_T}(u)\big]^2 W(u) {\rm{d}}u,  
\end{equation}
where $W: \mathbb R_+^d \to [0,\infty)$ is an integrable weight function whose choice is discussed later. 
Under the null hypothesis, the two functions $\widehat{L}_T(u)$ and $L_{\widehat{\varphi}_T}(u)$ will be close, while the opposite is true for the alternative. Consequently, 
large values of $S_{T,W}$ indicate violation of  the null hypothesis. 

\smallskip

Remark that the test statistic $S_{T,W}$ depends 
on the estimator $\widehat \vartheta_T$ of the vMEM parameter $\vartheta$ as well as on the estimator $\widehat\varphi_T$ of the distributional parameter $\varphi$.  One possibility is to construct the two estimators jointly via the maximum likelihood method as in  \cite{Cip17}.  This approach also includes a copula--based conditional distribution of $x_t$. 
Another approach is to proceed in two steps and estimate first $\widehat{\vartheta}_T$ using a  generalized method of moments (GMM) and consequently estimating $\varphi$ from the residuals $\{\widehat{\eps}_t\}_{t=1}^T$.

\subsection{Weight function and computation}\label{sec:comp}

Straightforward calculations lead to the following expression for the test statistic 
\[
S_{T,W} = \frac{1}{T}\sum_{t=1}^T\sum_{s=1}^T \xi_W(\what{\eps}_t+\what{\eps}_s) + T \eta_W(\widehat{\varphi}_T) - 2\sum_{t=1}^T \lambda_W(\widehat{\eps}_t;\widehat{\varphi}_T),
\]
where
\begin{align*}
\xi_W(x) &= \int_{\R^d_+} \mathrm{e}^{-u^\top x} W(u) \mathrm{d}{u},\qquad
\eta_W(\varphi)  =  \int_{\R^d_+} L_{\varphi}^2(u) W(u) \mathrm{d}{u},\\
\lambda_W(x;\varphi) & =   \int_{\R^d_+} \mathrm{e}^{-u^\top x} L_{\varphi}(u) W(u) \mathrm{d}{u}.
\end{align*}
Certain choices of the weight function  $W$ considerably simplify the above integrals. Namely,  let  $w_j$ be a  density of an absolutely continuous probability distribution on $\mathbb R_+$,  $j=1,\dots,d$, and  set 
$
W(u)= \prod_{j=1}^d w_j(u_j), \  u=(u_1,\dots,u_d)^\top. 
$
Then 
\begin{equation}\label{eq:xi}
\xi_W(x) = \int_{\R^d_+} \prod_{j=1}^d \mathrm{e}^{-u_jx_j} w_j(u_j) \mathrm{d}{u}_j = \prod_{j=1}^d \mathcal{L}_{w_j}(x_j),
\end{equation}
where  $\mathcal{L}_{w_j}(x_j)$ is the LT of $w_j$. Hence, it is suitable to select $w_j$  such that  $\mathcal{L}_{w_j}$ has a simple closed form. 
For instance, one gets
\begin{equation}\label{L:gamma}
{\cal{L}}_{w_j}(x)=(1+\gamma x)^{-\kappa}, 
\end{equation}
if $w_j$ is the density of the Gamma distribution with scale $\gamma>0$ and shape parameter $\kappa>0$.
Furthermore, we get by changing the order of integration and from \eqref{eq:xi} that
\begin{align*}
\lambda_W(x;\varphi)  &=  \int_{\R^d_+} \mathrm{e}^{-u^\top x} \int_{\R^d_+} \mathrm{e}^{-u^\top y} \mathrm{d}G_{\varphi}(y) W(u) \mathrm{d}{u} =  \int_{\R^d_+} \xi_W(x+y) \mathrm{d}G_{\varphi}(y)  \\
&=\Ee\Bigl\{\prod_{j=1}^d {\cal{L}}_{w_j}(Z_j+x_j)\Bigr\},
\end{align*}
where   $Z=(Z_1,\dots,Z_d)^\top$ is a random vector with distribution $G_{\varphi}$.

\subsection{Asymptotics} \label{sec_as}

The asymptotic distribution of $S_{T,W}$ is derived under the following set of assumptions. 
\begin{itemize} 
\setlength{\leftmargin}{3em}
\setlength{\labelwidth}{1em}
\item[A1] Let $\{x_t\}$ be a strictly stationary and ergodic solution of \eqref{vMEM} for the parameter $\vartheta_0$ and $\{\eps_t\}$ with distribution $F_{\eps}$. 
Assume that  $\vartheta_0$ belongs to a compact set $\Omega_0\subset \R^K$ and $\mu(\cdot;\cdot) = \bigl(\mu_1(\cdot;\cdot),\dots, \mu_d(\cdot;\cdot)\bigr)^\top$ satisfies 
$\mu_j(\vartheta;\cdot)\geq \omega_0$ for all $j=1,\dots,d$, all $\vartheta\in\Omega_0$ and some $\omega_0>0$. Furthermore, there exist 
$K>0$ and $\rho\in(0,1)$ such that $
\| \mu(\vartheta;\widetilde{\mathcal{F}}_t) - \mu(\vartheta;\mathcal{F}_t)\| < K \rho^t$  for all $t$ and all $\theta\in \Omega_0$. 
\item[A2]
Assume that 
$F_{\eps}=G_{\varphi_0}$ for   $\varphi_0\in\Phi_0\subset\Phi$. Assume that  $L_{\varphi}$ is twice continuously differentiable with respect to $\varphi$ on $\Phi_0$ and 
\[
\int_{\R_+^d} \sup_{\varphi \in\Phi_0}  \left|\frac{\partial L_{\varphi} (u)}{\partial \varphi_j} \right|^2 W(u) \mathrm{d} u <\infty, \quad  \int_{\R_+^d} \sup_{\varphi \in\Phi_0} \left|\frac{\partial^2 L_{\varphi} (u)}{\partial \varphi_j\partial \varphi_k}\right|^2 W(u) \mathrm{d} u <\infty
\]
for all $j,k=1,\dots,r$, where $\varphi=(\varphi_1,\dots,\varphi_r)^\top$. 
Furthermore, let $\frac{\partial^2 L_{\varphi} (u)}{\partial \varphi_j\partial u_i}$ be continuous for all $\varphi\in\Phi_0$ and $u\in\R^d_+$ and for all $j=1,\dots,r$, $i=1,\dots,d$. 
\item[A3] Let the estimators $\widehat{\vartheta}_T$ and $\widehat{\varphi}_T$ satisfy
\begin{align}
\sqrt{T}(\widehat{\vartheta}_T - \vartheta_0) &= \frac{1}{\sqrt{T}}\sum_{t=1}^T m_{1,t} + \op(1), \label{eq:vartheta}\\
\sqrt{T}(\widehat{\varphi}_T - \varphi_0) &= \frac{1}{\sqrt{T}}\sum_{t=1}^T m_{2,t} + \op(1), 
\end{align}
where $\{m_{t}\} = \{(m_{t,1},m_{t,2})^\top\}$ is a strictly stationary sequence of martingale differences with respect to $\{\mathcal{F}^{\eps}_{t-1}\}$, where $\mathcal{F}^{\eps}_{t-1} =\sigma\{\eps_s, \ s\leq t-1\}$, with a finite variance matrix. 
\item[A4]    
Assume that $\mu(\vartheta,\cdot)$ is twice  continuously differentiable with respect to $\vartheta$ on the interior of $\Omega_0$.
Let 
\[
\Ee\|x_t\|<\infty, \quad \Ee\left\{\Bigl|\frac{\partial \mu_j}{\partial \vartheta_k} (\vartheta_0;\mathcal{F}_{t-1}) \Bigr| \frac{\eps_{t,j}}{\mu_j (\vartheta_0;\mathcal{F}_{t-1})} \right\}<\infty,
\]
 for all $j=1,\dots,d$, $k,=1,\dots,K$.  Let there exist a neighbourhood $U(\vartheta_0)$ of $\vartheta_0$ such that
\begin{align*}
\Ee\left\{\|x_t\|^2 \sup_{\vartheta\in U(\vartheta_0)} \Bigl| \frac{\partial \mu_j(\vartheta;\mathcal{F}_{t-1})}{\partial \vartheta_k}\frac{\partial \mu_j(\vartheta;\mathcal{F}_{t-1})}{\partial \vartheta_l}\Bigr| \right\}&<\infty,\\
\E\left\{ |x_{t,j}| \sup_{\vartheta\in U(\vartheta_0)} \Bigl| \frac{\partial^2 \mu_j(\vartheta;\mathcal{F}_{t-1})}{\partial \vartheta_k \partial \vartheta_l}\Bigr| \right\}&<\infty,
\end{align*}
 for all $j=1,\dots,d$, $k,l,=1,\dots,K$.
\item[A5] Let $W:\R^d_{+}\to [0,\infty)$ be an integrable function such that  
$\int_{\R^d_+} \|u\|^4 W(u) \mathrm{d} u$ is finite. 

\end{itemize}

In the following,  $\mathrm{diag}\{a\}$ stands for a diagonal matrix with a vector $a$ on the main diagonal. Denote as
 \begin{equation}\label{eq:gamma}
\Gamma(u)= \Ee\left[\mathrm{e}^{-u^\top \eps_t} [\mathrm{diag}\{ \mu(\vartheta_0,\mathcal{F}_{t-1})\}]^{-1}  \cdot  \mathrm{diag}\{ \eps_t\} \cdot \nabla_{\vartheta} \mu_t(\vartheta_0)\right],
 \end{equation}
where $\nabla_{\vartheta} \mu_t(\vartheta) = \left(\partial \mu_{i}(\vartheta,\mathcal{F}_{t-1})/\partial \vartheta_j \right)_{i=1,j=1}^{d,K}$ is a $d\times K$  matrix of the first derivatives of $\mu$ with respect to $\vartheta$. Since $\mathrm{e}^{-u^\top \eps_t} \leq 1$, it follows from A4 that $\Gamma(u)$ is a finite $d\times K$ matrix. 

\begin{theorem} \label{th1} Under A1--A5, the test statistic $S_{T,W}$ converges in distribution to 
$
\int_{\R^d_+} Z(u)^2W(u)\mathrm{d} u,
$
as $T\to\infty$, 
where $\{Z(u), u \in \R^d_+\}$ is a centred Gaussian process with covariance function  
\begin{align*}
\Ee Z(u) Z(s) =& \Ee\left(\mathrm{e}^{-u^\top\eps_t} - L_{\varphi_0}(u) + u^\top \Gamma(u) m_{t,1} - \dot{L}_{\varphi_0}(u)^\top m_{t,2} \right)\\
&\times \left(\mathrm{e}^{-s^\top\eps_t} - L_{\varphi_0}(s) + s^\top\Gamma(s) m_{t,1}- \dot{L}_{\varphi_0}(s)^\top m_{t,2} \right),
\end{align*}
for $u,s\in \R^d_+$, where  $\dot{L}_{\varphi_0}(u) = \frac{\partial L_{\varphi}(u)}{\partial \varphi}\Big|_{\varphi=\varphi_0} $ and $m_{t,j}$, $j=1,2$, are from A3. 
\end{theorem}

Assumptions A1 and A2 specify that the null hypothesis holds with the requirement that  $L_{\varphi}(u)$ is a sufficiently smooth function of $\varphi$ and $u$. The representation of $\what{\vartheta}_T$ and  $\what{\varphi}_T$ required in A3 is rather a standard assumption. 
The inequality assumed in A1 for $\| \mu(\vartheta;\widetilde{\mathcal{F}}_t) - \mu(\vartheta;\mathcal{F}_t)\|$ can be typically shown for GARCH or vMEM processes under some additional technical assumptions on the function $\mu$ in \eqref{vMEM}, 
see, e.g., (4.6) in \cite{FZ} for a univariate GARCH and Lemma A.1 in \cite{Aknouche} for a univariate MEM.
We prefer to avoid these technical assumptions and so this property is directly formulated as an assumption of the theorem. Condition A4 requires the mean function $\mu$ to be sufficiently smooth and existence of some moments. For instance for a GARCH type model \eqref{GARCH} with $q=0$, A4 holds simply whenever $\E\|x_t\|^4<\infty$. 

\medskip

The next theorem shows that the test, which rejects the null hypothesis $\mathcal{H}_0$ for large values of $S_{T,W}$, is consistent. 

\begin{theorem} \label{th2} Let A1, A4 and A5 hold and let \eqref{eq:vartheta} hold for $\what{\vartheta}_T$.  Let $F_{\eps}\not \in \mathcal{G}$. Assume that $L_{\widehat{\varphi}_T} (u)\Pto L_{\varphi_A} (u)$ uniformly in $u\in\R^{d}_+$ for some $\varphi_A\in\Phi$. Then 
\[
\frac{S_{T,W}}{T} \Pto \int_{\R^d_+} \left| L_{\eps}(u) - L_{\varphi_A}(u) \right|^2 W(u) \mathrm{d} u,
\] 
where $L_{\eps} (u)= \Ee \mathrm{e}^{-u^\top\eps_1}$ is the LT of $\eps_t$.  
\end{theorem}

\subsection{Bootstrap test}\label{sec_boot1}

Since the asymptotic distribution of $S_{W,T}$ provided in Theorem~\ref{th1} depends on various unknown quantities in a complicated way, we suggest using a parametric bootstrap similar to the one suggested by \cite{Perera}. Let $\mathcal{B}$ be a selected number of bootstrap samples.  
\begin{enumerate}
\item For $\{x_t\}_{t=1}^T$ compute $\widehat{\vartheta}_T$, $\widehat{\varphi}_T$. Compute the test statistic $S_{T,W}$ from the residuals $\{\widehat\varepsilon_t\}_{t=1}^T$.
\item For $b=1,\dots,\mathcal{B}$:
\begin{enumerate}
\item Simulate bootstrap innovations $\{\eps_t^{(b)}\}_{t=1}^T$ as independent observations from the distribution $G_{\widehat{\varphi}_T}$. 
\item  Compute the bootstrap observations $\{x_t^{(b)}\}_{t=1}^T$ recursively from \eqref{vMEM} with $\vartheta$ replaced by $\widehat{\vartheta}_T$ and some initial values of $x_t$ and $\mu_t$.
\item Compute the estimators $\widehat{\vartheta}_T^{(b)}$, $\widehat{\varphi}_T^{(b)}$  and the residuals $\{\widehat{\eps}_t^{(b)}\}$ from the bootstrap series $\{x_t^{(b)}\}_{t=1}^T$. Calculate the corresponding test statistic $S_{T,W}^{(b)}$. 
\end{enumerate}
\item
The bootstrap estimate of the $p$-value is ${\mathcal{B}}^{-1}\sum_{b=1}^{\mathcal{B}} {\bf{1}}[  S_{T,W}^{(b)}> S_{T,W}]$. 
 \end{enumerate} 

 \begin{remark}
In the simulation study in Section~\ref{sec_6}, we simulate bootstrap series of length $T+m$ in steps 2.a.--b. and then the first $m$ observations are left out to get an approximately stationary series  $\{x_t^{(b)}\}_{t=1}^T$ used for the computation of the test statistic in step 2.c. We set $m=200$. 
\end{remark}

\section{Test statistic based on artificial sample}\label{sec_artif}

The test based on  $S_{T,W}$ requires the LT $L_\varphi$ to be known and reasonably simple so that integration over $\mathbb R_+^d$ can be manageable for arbitrary dimension $d\geq 1$. However, this is not the case for many multivariate distributions, where  $L_{\varphi}(u)$ is typically either completely unknown, or complicated for multidimensional integration.  In order to circumvent this drawback we define a test statistic that makes use of an artificial sample simulated from $G_{\widehat{\varphi}_T}$. 
Note that the idea of a goodness--of--fit method employing an artificial sample from the distribution under test seems to date back to \cite{friedman03}, at least for independent data and simple hypotheses. Recently \cite{CJMZ} proposed a similar method for goodness--of--fit within the family of multivariate elliptical distributions with estimated location and scatter matrix, while \cite{davis21} suggested an estimation method for dynamic data using the same notion.      

Let $S>1$ and $\{\varepsilon^*_t\}_{t=1}^S$ be independent  observations generated from $G_{\widehat{\varphi}_T}$. 
Consider
\begin{equation} \label{test}
\Psi_{T,W}=\frac{TS}{T+S} \int_{\mathbb R_+^d} \big[\widehat{L}_T(u)-\widehat L^*_{S}(u)\big]^2 W(u) {\rm{d}}u,  
\end{equation}
where
$\widehat L^*_{S}(u)= \frac{1}{S}\sum_{t=1}^S \mathrm{e}^{- u^\top \eps_{t}^*}$, 
is the empirical LT corresponding to the artificial sample.  Then the null hypothesis $\mathcal{H}_0$ should be rejected for large values of $\Psi_{T,W}$.

The choice $W(u)= \prod_{j=1}^d w_j(u_j)$ for $w_j$ being a density function for $j=1,\dots,d$, we get  
\begin{align*} \label{testsum}
\Psi_{T,W}&= \frac{S}{T(T+S)}\sum_{t,s=1}^T \prod_{j=1}^d {\cal{L}}_{w_j}(\widehat\varepsilon_{t,j}+\widehat\varepsilon_{s,j})
+ \frac{T}{S(T+S)}\sum_{t,s=1}^S \prod_{j=1}^d {\cal{L}}_{w_j}
(\varepsilon^*_{t,j}+\varepsilon^{*}_{s,j}) \\ 
\nonumber &-\frac{2}{T+S}\sum_{t=1}^T \sum_{s=1}^S\prod_{j=1}^d {\cal{L}}_{w_j}(\widehat\varepsilon_{t,j}+\varepsilon^{*}_{s,j}),
\end{align*}
where ${\cal{L}}_{w_j}$ is the LT of $w_j$.  Suitable choices of $w_j$ lead to closed form expressions, as, e.g., 
\eqref{L:gamma}. 
Hence, the computation of $\Psi_{T,W}$ is indeed easier than the computation of $S_{T,W}$, because a fully closed form expression is available for $\Psi_{T,W}$.

\subsection*{Bootstrap test}\label{sec_boot2}

It would be possible to formulate assertions analogous to Theorems~\ref{th1} and \ref{th2} about the asymptotic behavior of $\Psi_{T,W}$ as $T\to\infty$. 
 Like $S_{T,W}$, it is expected that the asymptotic null distribution
 of $\Psi_{T,W}$  could not  be easily used if $L_{\varphi}$ is unknown or complicated. Therefore, we again recommend a parametric bootstrap procedure. 

For fixed data $x_1,\dots,x_T$, the value of the test statistic 
$\Psi_{T,W}$, and consequently the decision of rejecting or not rejecting the null hypothesis $\mathcal{H}_0$, depends on the artificial sample $\{\eps_t^*\}_{t=1}^S$. Since this randomness can be seen as a drawback for practical usage, \cite{CJMZ} 
recommend, in an analogous situation, to generate $M$ samples instead of one and to compute the final test statistic as either the maximum or mean of the $M$ replicas. The bootstrap procedure then proceeds as follows:

\begin{enumerate}
\item For $\{x_t\}_{t=1}^T$ compute $\widehat{\vartheta}_T$, $\widehat{\varphi}_T$ and the residuals $\{\widehat\varepsilon_t\}_{t=1}^T$.
\item For $m=1,\dots,M$:
\begin{enumerate} 
\item Generate  independent realizations $\{\varepsilon_{t}^{*m}\}_{t=1}^T$ from 
$G_{\widehat \varphi_T}$.
\item Compute the test statistic $\Psi_{T,W}^{(m)}$ from  $\{\widehat\varepsilon_t\}_{t=1}^T$ and $\{\varepsilon_{t}^{*m}\}_{t=1}^T$.
\item Set the test statistic as the quantity
\begin{equation}\label{eq:meanPsi}
\overline{\Psi}_{T,W,M} = \frac{1}{M}\sum_{m=1}^M \Psi_{T,W}^{(m)}. 
\end{equation}
\end{enumerate}
\item For $b=1,\dots,\mathcal{B}$:
\begin{enumerate}
\item Simulate bootstrap innovations $\{\eps_t^{(b)}\}_{t=1}^T$ as independent observations from the distribution $G_{\widehat{\varphi}_T}$. 
\item Compute the bootstrap observations $\{x_t^{(b)}\}_{t=1}^T$ recursively from \eqref{vMEM} with $\vartheta$ replaced by $\widehat{\vartheta}_T$ and some initial values of $x_t$ and $\mu_t$.
\item Compute the estimators $\widehat{\vartheta}_T^{(b)}$, $\widehat{\varphi}_T^{(b)}$  from the bootstrap series and the residuals  $\{\widehat\varepsilon_t^{(b)}\}_{t=1}^T$.
\item For $m=1,\dots,M$,  generate  independent realizations $\{\varepsilon_{t}^{*(b,m)} \}_{t=1}^T$ from 
$G_{\widehat \varphi_T^{(b)}}$ and  compute  $\Psi_{T,W}^{(b,m)}$ from $\{\widehat\varepsilon_t^{(b)}\}_{t=1}^T$ and $\{\varepsilon_{t}^{*(b,m)} \}_{t=1}^T$. 
\item Take
\begin{equation}\label{eq:meanPsiB}
\overline{\Psi}_{T,W,M}^{(b)} = \frac{1}{M}\sum_{m=1}^M \Psi_{T,W}^{(b,m)}. 
\end{equation}
\end{enumerate}
\item The bootstrap estimate of the $p$-value is 
\[
\frac{1}{\mathcal{B}}\sum_{b=1}^{\mathcal{B}} {\bf{1}}[  \overline{\Psi}_{T,W,M}^{(b)}> \overline{\Psi}_{T,W,M}].
\] 
\end{enumerate} 
The choice of the number of artificial samples $M$ is numerically studied in Section~\ref{sec_6}, where we also explore a test with $\overline{\Psi}_{T,W}$ and $\overline{\Psi}_{T,W}^{(b)}$  in \eqref{eq:meanPsi} and \eqref{eq:meanPsiB} replaced by 
\[
\widetilde{\Psi}_{T,W,M} = \max_{1\leq m\leq M} \Psi_{T,W}^{(m)} \quad \text{ and } \quad \widetilde{\Psi}_{T,W,M}^{(b)} = \max_{1\leq m\leq M} \Psi_{T,W}^{(b,m)}. 
\]
If $M=1$, then $\overline{\Psi}_{T,W,M} =\widetilde{\Psi}_{T,W,M}$.

\section{Simulations} \label{sec_6}

The small sample behavior of the tests based on $S_{T,W}$ and $\overline{\Psi}_{T,W,M}$ is investigated in a simulation study for $d=2$
and data $\{x_t\}_{t=1}^T$ simulated from vMEM(1,1) process in~\eqref{GARCH} with 
\[
\alpha_0 = \begin{pmatrix} 0.2 \\ 0.2 \end{pmatrix}, \quad A_1 = A= \begin{pmatrix} 0.3 & 0.1 \\ 0.1& 0.3 \end{pmatrix}, \quad B_1 = B = \begin{pmatrix} 0.3 & 0.0 \\ 0.0 & 0.3 \end{pmatrix}.
\]
An interesting hypothesis to be tested is that the two components of $\eps_t$ are independent and follow a Gamma distribution, as this is the important special case of the multivariate distributions considered in \cite{Cip06}. Recall that it is required that $\Ee\eps_{t,j}=1$, $j=1,2$, and, therefore, there is one free parameter for each marginal Gamma distribution. Formally, we test
\eqref{null} with 
\[
 \mathcal{G} = \{G_{(\beta_1,\beta_2)}: \ G_{(\beta_1,\beta_2)} (x_1,x_2)= F_{\Gamma(\beta_1,\beta_1)}(x_1)\cdot F_{\Gamma(\beta_2,\beta_2)}(x_2)\},
\]
where $F_{\Gamma(\alpha,\lambda)}$ stands for the cumulative distribution function (CDF) of a Gamma distribution with shape $\alpha>0$ and rate $\lambda>0$ with mean $\alpha/\lambda$ and variance $\alpha/\lambda^2$. 
The weight function $W$ is chosen such that $W(u)=\prod_{j=1}^2 w(u_j)$ where $w$ is the density of Gamma distribution with shape $\kappa>0$ and scale $\gamma>0$ (i.e. the rate is $1/\gamma$), and $\mathcal{L}_w$ is given in \eqref{L:gamma}. The parameters $\vartheta=(\alpha_{0,1},\alpha_{0,2},A_{11},A_{12},A_{21},A_{22},B_{11},B_{22})^\top$ and $\varphi=(\beta_1,\beta_2)^\top$ are estimated jointly by the maximum likelihood method (MLE) knowing the diagonal structure of $B$.

\begin{table}[ht]
\caption{Size and power of the test based on the mean statistic $\widetilde{\Psi}_{T,W,M}$ and maximal statistic $\overline{\Psi}_{T,W,M}$ for $M\in\{1,5,10\}$ and $\kappa=\gamma=1$ under $\mathcal{H}_0$ and under $\mathcal{H}_1$  in \eqref{eq:alt} with $C_{\rho}$ being a Gaussian copula.}\label{tab:pilot}
\centering 
\begin{tabular}{rr|c|cc|cc}
\hline\noalign{\smallskip}
 & $T$ & $M=1$ & \multicolumn{2}{c|}{$M=5$} & \multicolumn{2}{c}{$M=10$} \\ 
 && & $\overline{\Psi}_{T,W,M}$& $\widetilde{\Psi}_{T,W,M}$&  $\overline{\Psi}_{T,W,M}$& $\widetilde{\Psi}_{T,W,M}$\\
\noalign{\smallskip}\svhline\noalign{\smallskip}
$\mathcal{H}_0$& 500 & 0.044 & 0.053 & 0.065 & 0.048 & 0.061 \\ 
  & 1000 & 0.070 & 0.097 & 0.068 & 0.050 & 0.054 \\ 
   & 2000 & 0.066 & 0.039 & 0.037 & 0.046 & 0.056 \\ 
   \hline
  Gaussian $\rho=0.10$  & 500 & 0.063 & 0.068 & 0.047 & 0.113 & 0.079 \\ 
  & 1000 & 0.068 & 0.139 & 0.095 & 0.185 & 0.113 \\ 
   & 2000 & 0.102 & 0.250 & 0.171 & 0.468 & 0.242 \\ 
  Gaussian $\rho=0.20$  & 500 & 0.134 & 0.266 & 0.153 & 0.368 & 0.174 \\ 
   & 1000 & 0.178 & 0.571 & 0.347 & 0.761 & 0.371 \\ 
   & 2000 & 0.417 & 0.913 & 0.664 & 0.990 & 0.799 \\ 
\noalign{\smallskip}\hline\noalign{\smallskip}
\end{tabular}
\end{table}

\begin{table}[ht]
\caption{Empirical size of the test based on $S_{T,W}$ and $\overline{\Psi}_{T,W,M}$ with $M=10$ for testing independent marginals with a Gamma distribution.}\label{tab:size}
\centering  
\renewcommand{\arraystretch}{1.2}
\begin{tabular}{r|ccc|ccc}
\hline\noalign{\smallskip}
  & \multicolumn{3}{c|}{$S_{T,W}$}& \multicolumn{3}{c}{$\overline{\Psi}_{T,W,M}$}\\
 $(\kappa,\gamma)$ & $T=500$& $T=1\,000$ & $T=2\,000$ &$T=500$& $T=1\,000$ & $T=2\,000$ \\ 
\noalign{\smallskip}\svhline\noalign{\smallskip}
  (0.5,0.5) & 0.058 & 0.046 & 0.054 & 0.041 & 0.055 & 0.048 \\ 
  (0.5,1.0) & 0.048 & 0.046 & 0.046 & 0.043 & 0.051 & 0.044 \\ 
  (0.5,2.0) & 0.047 & 0.037 & 0.049 & 0.052 & 0.045 & 0.038 \\ 
  (1.0,0.5) & 0.059 & 0.043 & 0.051 & 0.042 & 0.055 & 0.045 \\ 
  (1.0,1.0) & 0.053 & 0.037 & 0.053 & 0.048 & 0.050 & 0.046 \\ 
  (1.0,2.0) & 0.053 & 0.042 & 0.049 & 0.055 & 0.054 & 0.044 \\ 
  (2.0,0.5) & 0.060 & 0.042 & 0.052 & 0.063 & 0.049 & 0.048 \\ 
  (2.0,1.0) & 0.057 & 0.059 & 0.050 & 0.064 & 0.055 & 0.052 \\ 
  (2.0,2.0) & 0.043 & 0.046 & 0.044 & 0.063 & 0.063 & 0.060 \\ 
\noalign{\smallskip}\hline\noalign{\smallskip}
\end{tabular}
\end{table}

The considered sample size is $T\in\{500,1\,000,2\,000\}$. The tuning parameters $\kappa$ and $\gamma$ are taken from the set $\{0.5,1.0,2.0\}$, leading to 9 possible combinations. 
Under the null hypothesis, $\eps_t$ is simulated from $G_{(\beta_1,\beta_2)}$ with $\beta_1=2$ and $\beta_2=3$. Under the alternative, $\eps_t$ has a CDF
\begin{equation}\label{eq:alt}
F_{\eps}(x_1,x_2) = C_{\rho}\left( F_{\Gamma(\beta_1,\beta_1)}(x_1),  F_{\Gamma(\beta_2,\beta_2)}(x_2)\right),
\end{equation}
where $C_{\rho}$ is either a Gaussian copula with parameter $\rho \in \{0.10;0.20\}$ or a Clayton copula with parameter  $\rho \in \{0.15;0.30\}$. Note that these different choices for $\rho$ for Gaussian and Clayton copula lead to comparable 
 strength of dependence, measured by Kendall's tau (Kendall's tau between 0.06--0.07 for the smaller values of $\rho$, and Kendall's tau 0.13 for the larger values).

\begin{figure}[htbp]
\centering
\includegraphics[width=0.9\textwidth]{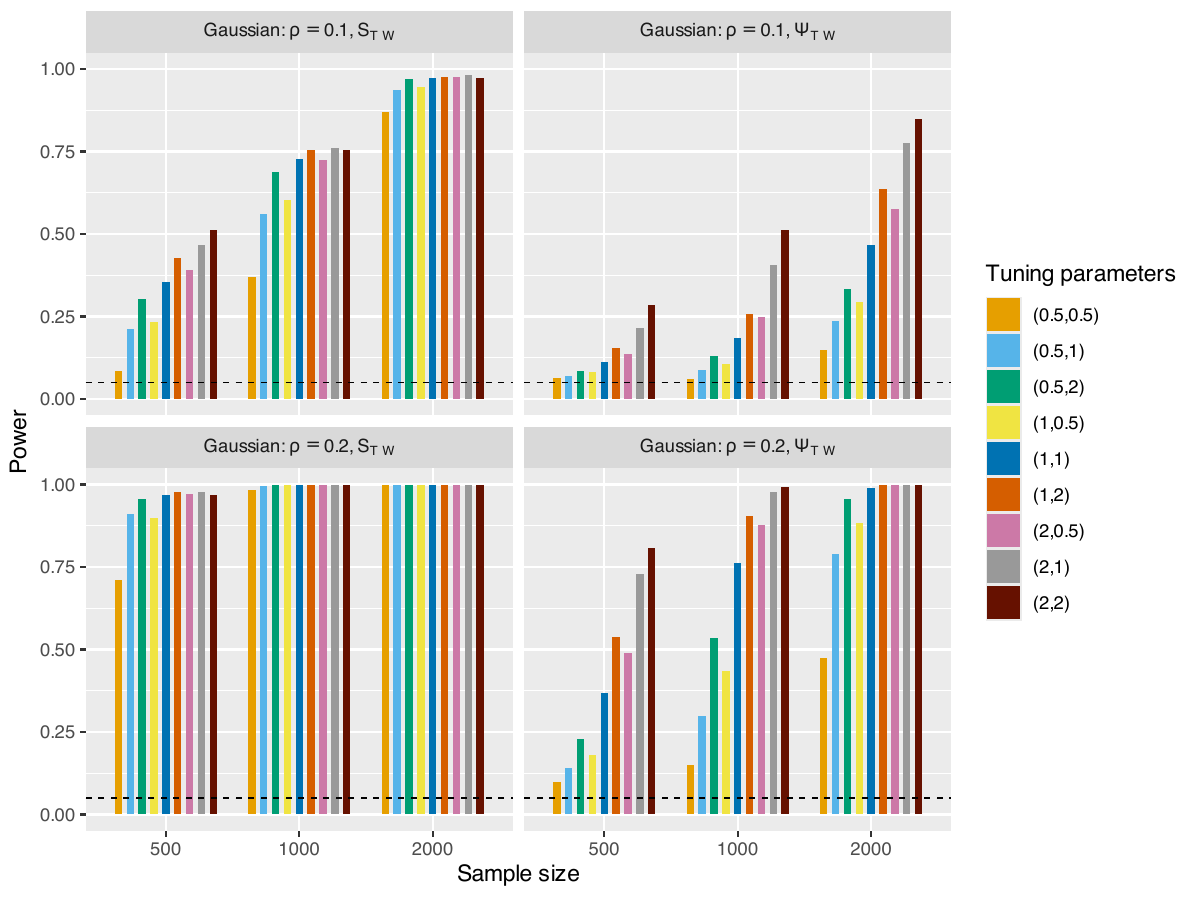}
\caption{Power for $S_{T,W}$ and $\overline{\Psi}_{T,W,M}$ with $M=10$ for data generated from \eqref{eq:alt} with a Gaussian copula $C_{\rho}$.}\label{fig:alt:Gaus}
\end{figure}

\begin{figure}[htbp]
\centering
\includegraphics[width=0.9\textwidth]{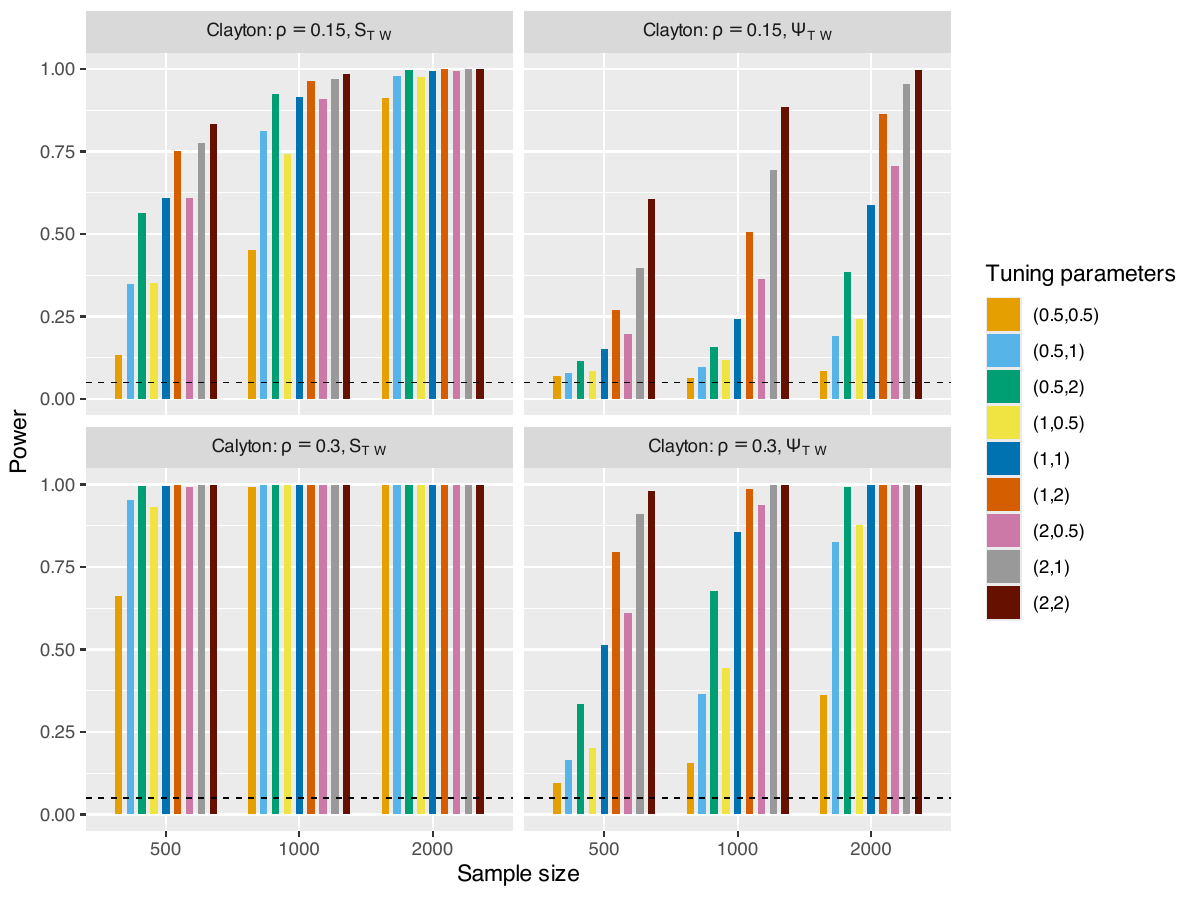}
\caption{Power for $S_{T,W}$ and $\overline{\Psi}_{T,W,M}$ with $M=10$ for data generated from \eqref{eq:alt} with a Clayton copula $C_{\rho}$.}\label{fig:alt:Clayton}
\end{figure}

  The empirical sizes and powers, provided in Tables~\ref{tab:pilot} and~\ref{tab:size} and Figures \ref{fig:alt:Gaus} and \ref{fig:alt:Clayton}, are computed from $1\,000$ simulated data sets using the warp-speed bootstrap of \cite{Giacomini}.
With this approach, there is solely one bootstrap sample for each Monte Carlo iteration, but  the significance is then evaluated from the whole set of $\mathcal{B}=1\,000$
   bootstrap samples.

 Table~\ref{tab:pilot} compares the performance of the test based based on the mean statistic $\overline{\Psi}_{T,W,M}$ and the maximal statistic $\widetilde{\Psi}_{T,W,M}$ for $M\in\{1,5,10\}$ under the null hypothesis and under the Gaussian alternative for tuning parameters $\kappa=\gamma=1.0$. All the variants lead to a similar empirical size, but the power increases with the number of artificial samples $M$.
 For a fixed value of $M$,  the mean statistic  $\overline{\Psi}_{T,W,M}$ achieves a larger power compared to the maximal statistic  $\widetilde{\Psi}_{T,W,M}$, so the largest power is  obtained for $\overline{\Psi}_{T,W,10}$.
 The same conclusion can be made for other choices of $(\kappa,\gamma)$ (results are not shown here for  the sake of brevity).   
 This finding is also in agreement with the results of \cite{CJMZ}. Therefore, we focus on $\overline{\Psi}_{T,W,M}$  with $M=10$ and present results solely for this setting.

 Table~\ref{tab:size} provides the empirical size of the tests based on $S_{T,W}$ and  $\overline{\Psi}_{T,W,M}$ under the null hypothesis for different $(\gamma,\kappa)$.
The test based on $S_{T,W}$ is well sized for all combinations of the tuning parameters, while $\Psi_{T,W}$ is very slightly oversized for $\gamma=\kappa=2.0$. The empirical power obtained  for data generated from \eqref{eq:alt} with Gaussian copula is provided in Figure~\ref{fig:alt:Gaus}.  The results show that the test statistic $S_{T,W}$ leads to a generally larger power compared to the test statistic  $\overline{\Psi}_{T,W,M}$ that makes use of $M$  artificial samples. The same conclusion is obtained for the Clayton copula in Figure~\ref{fig:alt:Clayton}. 
As expected, the power increases with the dependence parameter $\rho$.  
Comparing the tuning parameters, it seems that larger values for $\kappa$ and $\gamma$ are more suitable. Among the considered settings, the choice $\kappa=\gamma=2.0$ could be recommended. 

 
 \smallskip

The simulations suggest that when the  test statistic $S_{T,W}$ can be computed, then this method is preferred. The test based on $\overline{\Psi}_{T,W,M}$ should be used only if  the calculation of $S_{T,W}$ is not possible, and in that case, the choice $M\geq 10$ is recommended.

%

\section{Conclusion} \label{sec_7}

We propose a specification test for the error distribution in a vector multiplicative error model (vMEM). The test statistic is based on the weighted integral distance between two estimators of the Laplace transform (LT) of the error distribution: a nonparametric estimator and a parametric one. We show that the test is consistent, and that the bootstrap version performs well in finite samples. The power is evaluated under various settings of the tuning parameters.
If the LT of the null distribution is not available in closed form, the test is based on a two-sample comparison between the empirical LT of the residuals and the empirical LT of an artificial sample simulated from the null distribution with estimated parameters. To reduce randomness and increase power, it is advisable to take $M\geq 10$
 independent artificial samples and average the corresponding test statistics. However, even with this adjustment, this version of the test appears to be less powerful than the one that uses the parametric estimator of the LT.


%
%
%

\begin{acknowledgement}
The research of \v{S}\'{a}rka Hudecov\'{a} was supported by the Czech Science Foundation project GA\v{C}R No. 25-15844S. 
I am also deeply grateful for the long-term support of my dear colleague and former teacher,  Marie Hu\v{s}kov\'{a}, throughout my academic career. 

This paper was completed after the sad passing of Simos Meintanis. However, his contribution was fundamental, and I am grateful for the opportunity to have worked with him.   I would also like to emphasize that Simos was one of the first to propose the idea of a Festschrift for Professor Marie Hušková.

 Finally, I am also grateful to the reviewer for their valuable comments, which significantly improved the paper.
 \end{acknowledgement}

\ethics{Competing Interests}{
The authors have no conflicts of interest to declare that are relevant to the content of this chapter.}

%
%
%
%
%
%
%
%

\section*{Appendix: Proofs}\label{sec_proofs}
\addcontentsline{toc}{section}{Appendix}

Recall that $\widehat{\eps}_t =  x_t \oslash  \mu(\widehat{\vartheta}_T; \widetilde{\mathcal{F}}_{t-1})$ and 
define
$
\widetilde{\eps}_t = x_t \oslash  \mu(\widehat{\vartheta};\mathcal{F}_{t-1}),
$
the oracle residuals computed from the infinite past.
Furthermore, let 
\[
\widetilde{L}_T(u)  = \frac{1}{T} \sum_{t=1}^T \mathrm{e}^{-u^\top \widetilde{\eps}_t}, \quad L_T(u) =  \frac{1}{T} \sum_{t=1}^T \mathrm{e}^{-u^\top {\eps}_t}. 
\]
 All the convergences bellow are considered for $T\to\infty$.  The notations $\op(1)$ and $O_{\mathsf{P}}(1)$ have the standard meaning.

 \begin{lemma}\label{lem0}
Under A1, A3 and A4, 
\[
\widetilde{L}_T(u)  = L_T(u)+ u^\top\Gamma(u) (\what{\vartheta}_T-\vartheta_0) +B_T(u),
\]
where $\Gamma(u)$ is from \eqref{eq:gamma} and  $\sqrt{T} B_T(u) =\op(1)$ uniformly for $u\in F$ for any compact $F \subset \R^d$.
\end{lemma}
\begin{proof}
We have
\[
\widetilde{L}_T(u) = \frac{1}{T}\sum_{t=1}^T l_t(u,\widehat{\vartheta}), \quad {L}_T(u) = \frac{1}{T}\sum_{t=1}^T l_t(u,\vartheta_0),
\]
where $l_t(u,\vartheta) = \mathrm{e}^{-u^\top (x_t   \oslash \mu(\vartheta;\mathcal{F}_{t-1}))}.$ 
Taylor expansion gives
\[
\widetilde{L}_T(u)  = L_T(u) + A_{T,1}(u)^\top (\what{\vartheta}_T-\vartheta_0) + (\what{\vartheta}_T-\vartheta_0)^\top A_{T,2}(u) (\what{\vartheta}_T-\vartheta_0),
\]
where
\[
A_{T,1}(u) = \left(\frac{1}{T}\sum_{t=1}^T  \frac{\partial l_t}{\partial \vartheta} (u,\vartheta_0)\right), \quad A_{T,2}(u)  =  \left(\frac{1}{T}\sum_{t=1}^T \frac{\partial^2 l_t}{\partial \vartheta\partial \vartheta^\top} (u,\vartheta^*) \right),
\]
where $\vartheta^*$ is between $\vartheta_0$ and $\widehat{\vartheta}_T$.  
Furthermore, 
\begin{align*}
\frac{\partial l_t}{\partial \vartheta} (u,\vartheta_0) &= l_t(u,\vartheta_0)\cdot \nabla_{\vartheta} \mu_t(\vartheta_0)^\top \mathrm{diag}\{ \eps_t\} \cdot [\mathrm{diag}\{ \mu(\vartheta_0,\mathcal{F}_{t-1})\}]^{-1} u\\
&= \mathrm{e}^{-u^\top \eps_t} \cdot \nabla_{\vartheta} \mu_t(\vartheta_0)^\top \mathrm{diag}\{ \eps_t\} \cdot [\mathrm{diag}\{ \mu(\vartheta_0,\mathcal{F}_{t-1})\}]^{-1} u.
\end{align*}
 It follows from Assumption A4 and ergodicity in A1 that $ A_{T,1}(u)$ converges in probability to $\Ee\frac{\partial l_t}{\partial \vartheta} (u,\vartheta_0)$. 
Furthermore, it follows from A4 that $A_{T,2} (u)$ can be bounded by $\|u\|^2 \cdot O_{\mathsf{P}}(1)$. Together with A3,
\[
\sqrt{T} B_t(u) = \sqrt{T}(\what{\vartheta}-\vartheta_0)^\top A_{T,2}(u) (\what{\vartheta}-\vartheta_0) + \sqrt{T}(\what{\vartheta}-\vartheta_0)^\top\left[A_{T,1}(u) - \Ee A_{T,1}(u) \right] = \op(1).
\]
\end{proof}

\begin{lemma}\label{lem:R}
Under A1, 
\[
\| \widehat{\eps}_t - \widetilde{\eps}_t\| \leq \|x_{t}\|    \widetilde{K} \rho^t
\]
for a constant $\widetilde{K}>0$ and $\rho\in(0,1)$ from A1. 
\end{lemma}
\begin{proof}
Let $R_t  = \widehat{\eps}_t - \widetilde{\eps}_t$. For the $j$-th component,
\begin{align*}
R_{t,j}  &= \what{\eps}_{t,j}-\widetilde{\eps}_{t,j} =  x_{t,j} \left(\frac{1}{\mu_j(\what{\vartheta}_T;\widetilde{\mathcal{F}}_{t-1})} -  \frac{1}{\mu_j(\widehat{\vartheta}_T; \mathcal{F}_{t-1})} \right)\\ 
& =
 x_{t,j} \left(\frac{\mu_j(\what{\vartheta}_T; \mathcal{F}_{t-1}) - \mu_j(\what{\vartheta}_T;{\widetilde{\mathcal{F}}}_{t-1})}{\mu_j(\what{\vartheta}_T;\widetilde{\mathcal{F}}_{t-1})\mu_j(\widehat{\vartheta}_T; \mathcal{F}_{t-1})} \right).
\end{align*}
It follows from A1 that
$
|R_{t,j}| \leq \frac{x_{t,j}}{\omega_0^2} K \rho^t
$
and, therefore, 
$
\|R_t\| \leq \|x_{t}\|    \widetilde{K} \rho^t.
$
\end{proof}

\noindent{\it Proof of Theorem~\ref{th1}.}  
Define
\[
\widehat{Z}_T(u) = \sqrt{T}[\widehat{L}_T(u)-L_{\what{\varphi}_T}(u)],
\]
 so $S_{T,W} = \int_{\R^d_+} \widehat{Z}_T^2(u)W(u) \mathrm{d} u$. Similarly, set 
$\widetilde{Z}_T(u) = \sqrt{T}\left[\widetilde{L}_T(u) - L_{\widehat{\varphi}_T}(u) \right]$. 

Taylor expansion  implies that
\[
\widehat{Z}_{T}(u)-\widetilde{Z}_T(u)=\frac{1}{\sqrt{T}} \left(\sum_{t=1}^T - \mathrm{e}^{u^\top \eps_t} u^\top(\widehat{\eps}_t - \widetilde{\eps}_t) + \mathrm{e}^{-u^\top \eps_t^*}(\widehat{\eps}_t - \widetilde{\eps}_t)^\top u u^\top (\widehat{\eps}_t - \widetilde{\eps}_t) \right)
\]
for $\eps_t^*$ between $\what{\eps}_t$ and $\widetilde{\eps}_t$. It follows from Lemma~\ref{lem:R} that
\begin{align*}
|\widehat{Z}_{T}(u)-\widetilde{Z}_T(u)|&\leq \frac{1}{\sqrt{T}} \sum_{t=1}^T \|u\| \cdot \|x_t\| \widetilde{K} \rho^t  + \frac{1}{\sqrt{T}}\sum_{t=1}^T \|u\|^2  \cdot \|x_t\|^2 \widetilde{K}^2 \rho^{2t}.  
\end{align*}
Hence, the assumption $\Ee\| x_t\|<\infty$  in A4 and Lemma S1 in \cite{HH} imply that   $|\widehat{Z}_{T}(u)-\widetilde{Z}_T(u)| =(\|u\|+ \|u\|^2) \op(1)$. 

It follows from Lemma~\ref{lem0} that 
\begin{align*}
\widetilde{Z}_T(u)&=\sqrt{T}\left[ \widetilde{L}_T(u) - L_T(u)+L_T(u) - L_{\varphi_0}(u) + L_{\varphi_0}(u) - L_{\widehat{\varphi}_T}(u) \right]\\
& = \sqrt{T}\left[ L_t(u) - L_{\varphi_0}(u) \right]+ \sqrt{T} u^\top \Gamma(u) (\what{\vartheta}_T-\vartheta_0) + \op(1) + \sqrt{T}\bigl[  L_{\varphi_0}(u) - L_{\widehat{\varphi}_T}(u) \bigr].
\end{align*}
 Taylor expansion gives
\[
\sqrt{T}\bigl[L_{\varphi_0}(u) - L_{\widehat{\varphi}_T}(u) \bigr] =   - \sqrt{T}  \dot{L}_{\varphi_0}(u)^\top (\widehat{\varphi}_T-\varphi_0) +R_T(u),
\]
where $\dot{L}_{\varphi_0}(u) = \frac{\partial L_{\varphi}(u)}{\partial \varphi}\Big|_{\varphi=\varphi_0} $,   
$
R_T(u) =\frac12 \sqrt{T}(\widehat{\varphi}_T-\varphi_0)^\top \mathbb{L}(u,\varphi^*)   (\widehat{\varphi}_T-\varphi_0),  
$
for $\mathbb{L}(u,\varphi) = \left(\frac{\partial^2 L_{\varphi}(u)}{\partial \varphi_i \varphi_j} \right)_{i,j=1}^r$, where $\varphi^*$ lies between $\widehat{\varphi}_T$ and $\varphi_0$.  Under assumptions A2, A3 and A5, $ \int_{\R^d_+} [R_T(u)]^2 W(u) \mathrm{d} u = \op(1)$. 
Furthermore,  it follows from A3  that 
\begin{align*}
\sqrt{T} u^\top \Gamma(u) (\what{\vartheta}_T-\vartheta_0) &= u^\top \Gamma(u)  \frac{1}{\sqrt{T}}\sum_{t=1}^T m_{t,1} + u^\top \Gamma(u) \op(1),\\ 
\sqrt{T}  \dot{L}_{\varphi_0}(u)^\top (\widehat{\varphi}_T-\varphi_0) &= \frac{1}{\sqrt{T}}\sum_{t=1}^T \dot{L}_{\varphi_0}(u)^\top m_{t,2} +  \dot{L}_{\varphi_0}(u)^\top\op(1). 
\end{align*}

We get from the previous and from assumptions A2 and A5 that 
\[
\widehat{S}_{T,W} = \int_{\R^d_+} {Z}_T(u)^2 W(u) \mathrm{d} u +\op(1),
\]
where
\[
{Z}_T(u) =  \frac{1}{\sqrt{T}} \sum_{t=1}^T \left(\mathrm{e}^{-u^\top\eps_t} - L_{\varphi_0}(u) +u^\top \Gamma(u) m_{t,1}- \dot{L}_{\varphi_0}(u)^\top m_{t,2} \right).  
\]
Since $\{\eps_t\}$ are non-negative independent and identically distributed and $\{m_t\}$ is a strictly stationary sequence of martingale differences with a finite variance matrix,  ${Z}_T(u)$ is a properly normalized sum of   martingale differences
\[
\delta_t(u) = \mathrm{e}^{-u^\top\eps_t} - L_{\varphi_0}(u) +u^\top \Gamma(u) m_{t,1}-\dot{L}_{\varphi_0}(u)^\top m_{t,2} 
\]
with a finite variance matrix. Consequently,  the finite dimensional distribution of the process $\{Z_T(u), \ u\in\R^d_+\}$ converge to the finite dimensional distribution of $\{Z(u), u\in\R^d_+\}$.

Let  $F\subset \R^d_+$ be any compact set. We will show that $\int_F Z_T(u)^2 W(u) \mathrm{d}u$ converges to $\int_F Z(u)^2 W(u) \mathrm{d}u$. 
In view of Theorem~22 from \cite{ibragimov} and the consecutive remark, it suffices to show that (i) $\sup_{T,u\in F} \Ee Z_T(u)^2 <\infty$ and (ii) $\Ee |Z_T(u)-Z_T(s)|^2 \leq H \|u-s\|^{\alpha}$ for some $H,\alpha>0$ and for all $u,s\in F$.

For (i),  see that $\mathrm{e}^{-u^\top\eps_t}\leq 1$ and $L_{\varphi_0}(u) = \Ee \mathrm{e}^{-u^\top \eps_t} \leq 1$, so it follows from the stationarity and martingale difference property that
\[
\Ee Z_T(u)^2  = \Ee \delta_t(u)^2 \leq 3\big(4 + \|\Gamma(u)^\top u\|^2  \Ee \|m_{t,1}\|^2+  \| \dot{L}_{\varphi_0}(u)\|^2 \Ee \| m_{t,2}\|^2\big). 
\]
Consequently, (i) follows from A2, A3 and A4. 
Similarly,
\begin{align*}
\E| Z_T(u) &- Z_T(s)|^2 = \Ee |\delta_1(u)- \delta_1(s)|^2 \\
&\leq 3\big\{\Ee |f_1(u) - f_1(s)|^2 + \Ee |m_{t,1}^\top(\Gamma(u)^\top u - \Gamma(s) s)|+ \Ee |m_t^\top( \dot{L}_{\varphi_0}(u)- \dot{L}_{\varphi_0}(s) )| \big\},
\end{align*}
where $f_1(u) = \mathrm{e}^{-u^\top \eps_1} - L_{\varphi_0}(u)$. Hence, the property (ii) follows from the mean value theorem and from the fact that the functions $f_1(u)$,  $\dot{L}_{\varphi_0}(u)$, and $u^\top \Gamma(u)$ have continuous derivatives with respect to $u$. %

It follows from A2 and A5 that for any $\kappa>0$ there exists a compact $F_{\kappa}$ such that 
\[
\Ee \int_{\R^d_+\setminus F_{\kappa}} Z_T(u)^2  W(u) \mathrm{d} u \leq \kappa. 
\] 
Analogously, it is possible to show that  $\Ee \int_{\R^d_+\setminus F_{\kappa}} Z(u)^2  W(u) \mathrm{d} u \leq \kappa.$  This proves that $\int_{\R^d_{+}} Z_T(u)^2 W(u) \mathrm{d}u$ converges to $\int_{\R^d_{+}}  Z(u)^2 W(u) \mathrm{d}u$.
\qed

\medskip

\noindent{\it Proofs of Theorem~\ref{th2}.}
The proof proceeds along similar lines as the proof of Theorem~\ref{th1}, because 
 $S_{T,W}/T = \int_{\R^d_{+}} [\sqrt{T}^{-1}\what{Z}_T(u)]^2 W(u) \mathrm{d} u$. Under A1, A3 for $\what{\vartheta}_T$ and A4, it follows from Lemmas~\ref{lem0} and \ref{lem:R} that 
 \[
 \frac{1}{\sqrt{T}}\what{Z}_T(u) =\frac{1}{T}\sum_{t=1}^T[\mathrm{e}^{-u^\top\eps_t} - L_{\what{\varphi}_T}(u)] +  R_T(u),
 \]
 where $|R_T(u)|\leq C \|u\|^2 \op(1)/\sqrt{T} + \|\Gamma(u)^\top u\| (\widehat{\vartheta}_T-\vartheta_0)$ for a constant $C>0$. 
 The rest follows from A5,  the law of large numbers for the sequence $\{\eps_t\}$ and the assumption about limiting behavior of $L_{\what{\varphi}_T}(u)$. \qed

 \bibliographystyle{spbasic}
\bibliography{vmemLIT}

\end{document}